\documentclass[a4paper,noinfoline]{imsart}

\usepackage[paperheight=11in,paperwidth=8.5in]{geometry}
\usepackage[colorlinks,citecolor=blue,urlcolor=blue]{hyperref}

\usepackage{amsmath}
\usepackage{amssymb}
\usepackage{amsthm}
\usepackage{float}
\usepackage{graphicx}
\usepackage{multirow}
\usepackage[sort,square,numbers]{natbib}

\allowdisplaybreaks

\startlocaldefs
\newcommand{\blmd}{\bar{\lambda}}
\newcommand{\braces}{\set}
\newcommand{\brackets}[1]{\left(#1\right)}
\newcommand{\E}[1]{\mathbb{E}\left[#1\right]}
\newcommand{\I}[1]{\mathbb{I}\set{#1}}
\newcommand{\iid}{i.i.d.}
\newcommand{\inv}{^{-1}}
\newcommand{\mas}{MaSEPTiDE}
\newcommand{\rightbar}[1]{\left.#1\right|}
\newcommand{\rmd}{\,\mathrm{d}}
\newcommand{\sei}{SEISMIC}
\newcommand{\set}[1]{\left\{#1\right\}}
\newcommand{\smallE}[1]{\mathbb{E}[#1]}
\newcommand{\smallrightbar}[1]{#1\mid}
\newcommand{\T}{\tilde{T}}
\newcommand{\tid}{TiDeH}
\newcommand{\tN}{\tilde{N}}
\newcommand{\tp}{\tilde{p}}
\newcommand{\tr}{^{\top}}

\endlocaldefs

\graphicspath{{figures/}}

\begin{document}

\begin{frontmatter}

\title{Marked Self-Exciting Point Process Modelling 
       of Information Diffusion on Twitter}
\runtitle{\mas: A Model of Information Diffusion on Twitter}

\begin{aug}
\ead[label=e1]{feng.chen@unsw.edu.au}
\ead[label=e2]{waihong.tan@student.unsw.edu.au}
\runauthor{F. CHEN AND W.H. TAN}
\affiliation{UNSW Sydney}
\vspace{2.5mm}
\address{Feng Chen  \\
School of Mathematics and Statistics  \\
UNSW Sydney, NSW 2052, Australia \\
\printead{e1}}
\address{Wai Hong Tan \\
School of Mathematics and Statistics \\
UNSW Sydney, NSW 2052, Australia \\
\printead{e2}}
\vspace{2.5mm}
\end{aug}

\begin{abstract}
\label{SEC:abs}
Information diffusion occurs on microblogging platforms like Twitter
as retweet cascades.  When a tweet is posted, it may be retweeted and
henceforth further retweeted, and the retweeting process continues
iteratively and indefinitely.  A natural measure of the popularity of a
tweet is the number of retweets it generates.  Accurate predictions of
tweet popularity can assist Twitter to rank contents more effectively
and facilitate the assessment of potential for marketing and
campaigning strategies.  In this paper, we propose a model called the
Marked Self-Exciting Process with Time-Dependent Excitation Function,
or \mas\ for short, to model the retweeting dynamics and to predict
the tweet popularity.  Our model does not require expensive feature
engineering but is capable of leveraging the observed dynamics to
accurately predict the future evolution of retweet cascades.  We
apply our proposed methodology on a large amount of Twitter data and
report substantial improvement in prediction performance over existing
approaches in the literature.
\end{abstract}

\begin{keyword}
  \kwd{B-spline}
  \kwd{forecast}
  \kwd{Hawkes process}
  \kwd{integral equation}
  \kwd{nonstationary self-exciting point process}
  \kwd{popularity prediction}
  \kwd{simulation}
\end{keyword}

\end{frontmatter}

\section{Introduction}
\label{SEC:int}
The advancement of technology has dramatically changed the ways people
connect to each other over the past few years.  This contributes to
the increasing popularity of microblogging platforms, which integrate
the features of instant messaging and blogging, enabling users to
conveniently share contents like short sentences, images or videos.
Twitter is a microblogging service that allows the users to share
information in the form of 140-character messages called {\em tweets}.
As a tweet is posted by a user, it may be shared
by the {\em followers} of the original poster through
an action known as {\em retweeting}, which explicitly refers to the
original tweet via its unique identification number, or a {\em retweet}.
This retweeting process can iterate indefinitely, resulting in a cascade
of retweets.

Information diffusion modelling in Twitter has been an active field of
research.  \cite{Tumasjan2010} considered Twitter as a platform used
for political deliberation and analyzed tweet sentiments by machine
learning to forecast the results of elections.  By learning the
features, one can also find the likelihood of retweets based on the
interestingness of contents \citep{Naveed2011}. \cite{Matsubara2012}
proposed a model to account for the rise and fall of influence
propagation whereas \cite{Alves2016} recently modelled random series
of events prevalent in Twitter by Poissonian and self-feeding
processes.  These studies, however, do not emphasize on popularity
prediction, which is our primary concern.

The {\em popularity} of a tweet is naturally measured by the size
of the retweet cascade, or the number of retweets it generates.
The predictions of tweet popularity are important as they can
assist Twitter to rank contents more effectively and facilitate the
assessment of potential for marketing and campaigning strategies.  As
such, many models have been proposed to capture the retweeting
dynamics and to predict the popularity.  One noticeable work is by
\cite{Zaman2014}, who proposed a Bayesian approach which predicts
tweet popularity using network information.  Models based on the
theory of point processes \citep{Zhao2015,Kobayashi2016}, which do not
require network information, were also shown to have good prediction
performances.  The model proposed by \cite{Mishra2016}, on the other
hand, combines the point process models with feature based approaches
to predict the popularity.  Other models like the growth-adoption
model by \cite{Lymperopoulos2016}, the spatial-temporal heterogeneous
Bass model by \cite{Yan2016}, and the concept drift model by
\cite{Li2016}, were all proposed for the purpose of popularity
predictions.

Some models employed on other online social networks (OSNs) with the
same purpose of popularity predictions are also relevant as the
proposed methodologies may be applicable to predictions on Twitter
network.  Notably, \cite{Agarwal2009} proposed a dynamic linear
regression model to predict the click-through rate for Today Module on
Yahoo! Front Page.  Activities on platforms like Youtube and Digg were
also modelled, for instance using linear regression models
\citep{Szabo2010} and classification models \citep{Ahmed2013}.  Other
related works include the reinforced Poisson model \citep{Gao2015}
applied on Sina Weibo and the model by \cite{Wu2016} that incorporates
temporality and seasonality, applied on Flickr image data.

Recently \cite{Zhao2015} proposed a model termed the \sei\
(Self-Exciting Model of Information Cascades) to model the retweeting
dynamics on Twitter.  The model describes the retweet intensity of a
tweet, or the expected number of retweets per unit time, as a product
of the infectivity of the original tweet and the accumulated
excitation effect of all previous retweets.  \cite{Zhao2015}
estimated the infectivity as a function of time using a kernel
smoothing estimator, and the excitation function, or {\em memory
kernel}, using a graphical approach under the assumption that some
retweeting processes follow an inhomogeneous Poisson process with the
excitation function as its intensity function.
They also proposed to predict the future popularity of a tweet
based on calculating the expected number of future retweets by first
assuming that the infectivity remains constant since the censoring time,
and a subsequent ad hoc adjustment to the expectation to incorporate the
decaying trend of the infectivity.  They reported that the predictions
of tweet popularity using their approach outperformed those based on
competing approaches, under several performance
measures.  \cite{Kobayashi2016} proposed another model termed the \tid\
(Time-Dependent Hawkes) model, which models the retweet intensity
similar to the \sei, and estimated the infectivity and memory
kernel using similar nonparametric kernel smoothing
estimators.  \cite{Kobayashi2016} fitted a circadian rhythmic function
to the nonparametrically estimated infectivity function up to the
censoring time, and extrapolated it beyond the censoring time to
predict the number of future retweets.  With certain choices of the
smoothing parameters, the tweet popularity predictions based on the
\tid\ model are superior to those based on the \sei, especially on
long cascades.  However, \citeauthor{Kobayashi2016}'s approach requires
sufficiently long observation time on a retweet sequence to have
reliable estimation of the infectivity function, and the prediction
performance depends critically on the window size parameter used in
the estimation of the infectivity function.

In this work, we propose a marked self-exciting point process model to
capture the retweeting dynamics and to predict tweet popularity.
Our model is motivated by the \sei\ and the \tid\ model, and bears some
similarities to them.  However, our model has some important
advantages.  First, the intensity process in our model has a
linear form similar to that of the original self-exciting process
of \cite{Hawkes1971}, and therefore the resulting point process is
interpretable as a cluster Poisson process,
which means our model can be simulated using a cascading algorithm
similar to that used for the efficient simulation of Hawkes processes.
Second, the estimation of our model and the assessment of the
goodness-of-fit can be implemented using principled approaches from
point process theory, and the predictions based on our model can also
be done properly by exploiting the probabilistic properties of the
model, without resorting to ad hoc assumptions such as those needed by
the \sei.  Moreover, our model is found to be able to capture the
retweeting dynamics and make accurate popularity predictions based on
much shorter observation times than those required by the \tid\ model.

The rest of the paper is organized as follows.  In
Section~\ref{SEC:md} we describe the tweet data and previous models
for tweet popularity predictions which motivated our work, and show how
our model is built, how we estimate the parameters and evaluate the
goodness-of-fit, and finally how to make predictions.  In
Section~\ref{SEC:res} we apply the proposed model to the tweet data
and compare the prediction performance of our model with that of the
\sei\ and the \tid\ model, to show the superior performance of our
model.  Finally, in Section~\ref{SEC:con}, we conclude the paper with a
discussion.

\section{The data, the model, and the methodology}
\label{SEC:md}

\subsection{Twitter data}
\label{sec:data}
\begin{figure}[H]
   \centering
   \includegraphics[width=0.6\textwidth,height=0.6\textwidth,
                    keepaspectratio]{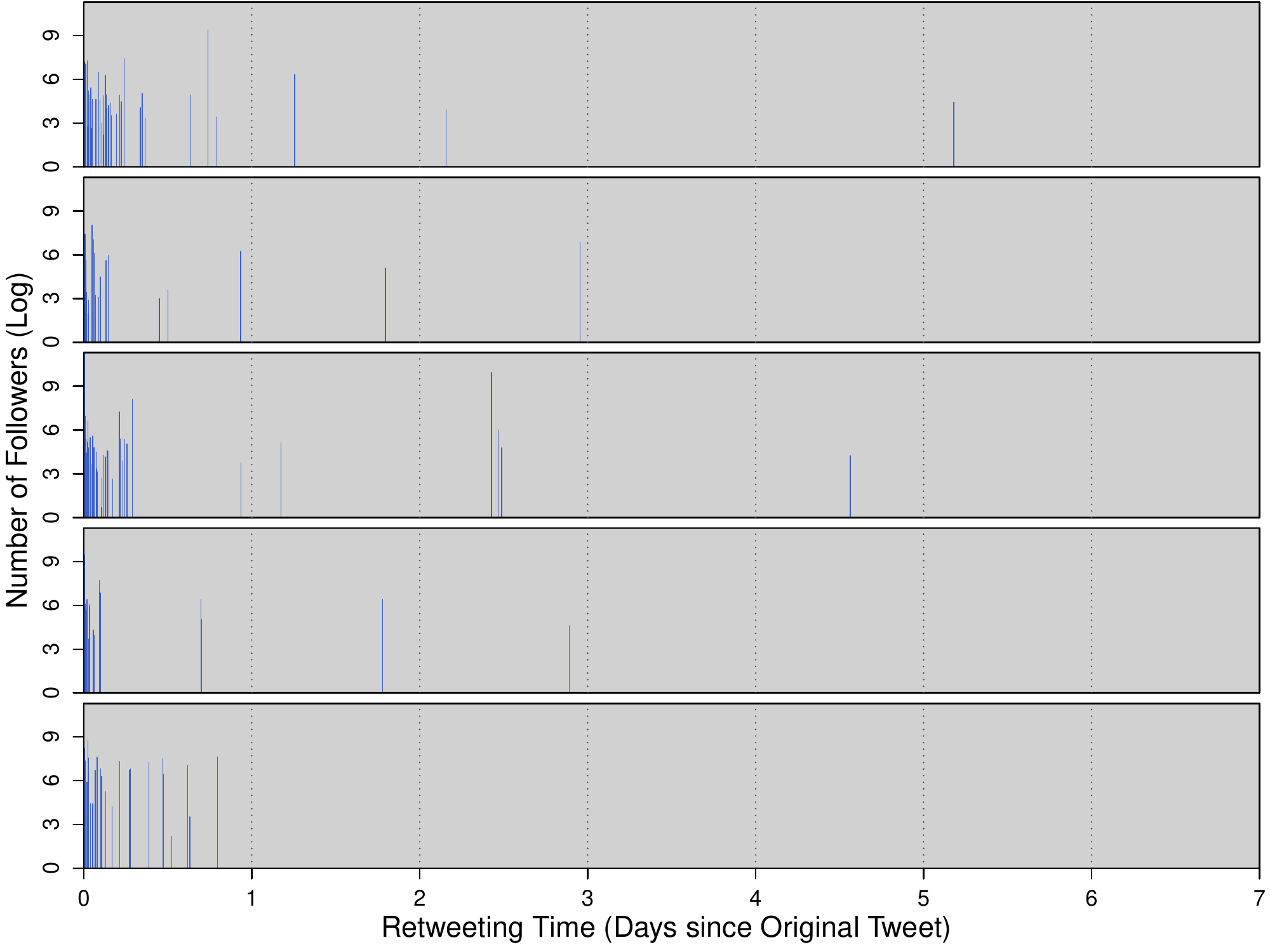}
   \caption{Times of retweets and the corresponding numbers of
            followers of the retweeting accounts on the log scale,
            for five randomly selected retweet cascades from the
            training data set.}
   \label{fig:smpdat}
\end{figure}

The data\footnote[1]{available from http://snap.stanford.edu/seismic/}
which motivated our work and which will be used to demonstrate our
modelling and prediction methodology is that used recently by
\cite{Zhao2015}.  The data contains a total of 166,069 reasonably
popular tweets published from October 7 to November 7, 2011, each with
at least 49 retweets within seven days of publishing.
For each tweet, the data includes the Twitter ID of the original
tweeter, the posting times of the original tweet and all the
retweets within seven days, and the numbers of followers of the
original poster and of the retweeters.  Following \cite{Zhao2015},
we use data on the 71,815 tweets published in the first seven days
of the study period as training data and the remaining 94,254
tweets published in the next eight days as test data.
See Figure~\ref{fig:smpdat} for 5 randomly selected
retweet cascades from the training data set.

Note that the data lacks the complete Twitter network information,
that is, for a retweet, the data only has its publishing time and the
number of followers of the retweeting account, without information on
whether the original tweet or any previous retweet is being retweeted.
This implies that the methodology of \cite{Zaman2014}, which assumes
the complete Twitter network information, does not apply here.  The
contents of the original tweets and of the retweets are also not
included in the data, and therefore methodologies depending on
features of the tweet contents or features of the posters
(other than their numbers of followers) do not apply either.

From Figure~\ref{fig:smpdat} we observe that the retweets tend to
occur in clusters or bursts.  This suggests that self-exciting
processes like the \sei\ of \cite{Zhao2015} are potentially
suitable for such data.  In the next subsection we present our model
for the retweeting dynamics, which is a marked self-exciting process
model similar to that of the \sei.

\subsection{Model formulation}
\label{sec:mdform}
Let $(\tau_i,m_i), i=0,1,\dotsc$ be a marked point process where
$0 = \tau_0<\tau_1<\dotsc$ denote the {\em event times} and
$m_0,m_1,\dotsc$ denote the respective {\em event marks}.  In the
context of information diffusion modelling on Twitter, the
event times shall refer to the retweeting times, except that
$\tau_0 = 0$ denotes the posting time of the original tweet, and
the event marks refer to the numbers of followers of the retweeting
(or tweeting, in the case of $m_0$) accounts.  Let
$N(t)=\sum_{i=1}^\infty \I{\tau_i\leqslant t},t\geqslant 0$
be the associated counting process of retweets, and
$\mathcal F=\set{\mathcal F_t; t\geqslant 0}$, with $\mathcal
F_t=\sigma\set{N(t), m_0, (\tau_j,m_j), j=1,\dotsc,N(t)}$, be the
natural filtration of the marked point process.  The
{\em (conditional) intensity process} of $N$ relative to the
filtration $\mathcal F$ is an $\mathcal F$-predictable process
$\lambda(t), t\geqslant 0$, such that $M(t)=N(t)-\int_0^t
\lambda(s)\rmd s, t\geqslant 0$ is a mean zero $\mathcal
F$-martingale.  In an informal but intuitive notation, the intensity
can be written as $\lambda(t)={\E{\rightbar{\rmd N(t)}\mathcal
F_{t-}}}/{\rmd t}$, from which we note that the intensity at any
time point is the expected number of events per unit time given the
history of the process prior to that time point.

As the evolution of a point process over time is fully determined by
its intensity process, a commonly used approach to specify a point
process model is to specify the form of the dependence of its
intensity process on the prior-$t$ history of the process $\mathcal
F_{t-}$.  In particular, the \sei\ of \cite{Zhao2015} assumes that
the intensity of the retweeting process $N(t)$ takes this form,
\begin{equation}
  \label{eq:lmd.sei}
  \lambda(t) = p(t)\sum_{i=0}^{N(t-)} m_i \phi(t-\tau_i),\quad t>0,
\end{equation}
where $p(t)$ is an unspecified positive function, called the
infectivity function, which typically decreases in $t$, and
$\phi(\cdot)$ is a positive function called the memory
kernel.  \cite{Zhao2015} proposed to estimate the infectivity function 
$p(\cdot)$ nonparametrically using a kernel smoothing estimator with a
triangular kernel.  To estimate the memory kernel, they assumed that it
is of a power law decaying form and that 15 ``carefully chosen''
retweet cascades follow inhomogeneous Poisson processes with
intensity functions proportional to the memory kernel.  They then
estimated the parameters using histogram and complementary cumulative
distribution function plots of the retweeting times in those 15
cascades.  As for the \tid\ model of \cite{Kobayashi2016}, it assumed
an intensity process of the same form as \eqref{eq:lmd.sei}, except
the further assumption that the infectivity function $p(\cdot)$ is
also parametric, and takes a dampened circadian oscillation form.  To
estimate the infectivity function $p(\cdot)$, \cite{Kobayashi2016}
proposed a two-step approach where a preliminary estimate $\hat
p(\cdot)$ was obtained first using a kernel method and then the
parametric form of $p(\cdot)$ was fitted to the preliminary estimate
by a least squares method.

The point process model we propose in the current work for the purpose
of retweeting dynamics modelling is given by
\begin{equation}
  \label{eq:lmd}
  \lambda(t)=\nu(t) + \sum_{i=1}^{N(t-)} \omega(\tau_i,m_i,t-\tau_i),
\end{equation}
where $\nu(\cdot)$ is the {\em baseline intensity function}, with
$\nu(t)$ denoting the part of the event intensity at time $t$ that is
due to the initial event at time $0$; and $\omega(\cdot,\cdot,\cdot)$
is the {\em excitation function},
with $\omega(\tau,m,t-\tau)$ denoting the
impact of an event at time $\tau$ with mark $m$ on the event intensity at
time $t$, where $t$ is the time since the initial tweet was posted.
Furthermore, both the baseline intensity
and the excitation functions are time-dependent and take
multiplicatively separable forms as follows,
\begin{equation}
  \label{eq:fn.sep}
  \begin{split}
    &\nu(t) = \alpha \phi(t),\\
    &\omega(\tau,m,t-\tau) = p(\tau)r(m)\phi(t-\tau).
  \end{split}
\end{equation}
Here $\alpha>0$ is a constant giving the direct excitation effect of
the original tweet, that is, how many retweets it is expected to generate
directly.  The function $\phi(\cdot)$ is called the {\em memory kernel
function}, which describes how the excitation effect due to the
original tweet or a retweet is distributed over time.  Similar to
\cite{Zhao2015}, we require $\phi(\cdot)$ to be a probability density
function, so that $\phi(\cdot)\geqslant 0$ and $\int_0^\infty
\phi(t)\rmd t=1$.  The function $p(\cdot)$ indicates how the
``infectivity'' of a retweet varies over time and is also called the
{\em infectivity function}, although its influence on the intensity
process $\lambda(t)$ is different than that of the infectivity
function $p(\cdot)$ in \eqref{eq:lmd.sei}.  For
identifiability, we assume that $p(0)=1$.  The function $r(\cdot)$ is
called the {\em impact function}, and describes the total excitation
effect of a retweet attributed to the number of followers
of the retweeter.  Note, we do not require $\alpha=r(m_0)$, to
allow for the potentially different influences of the original tweet
and of the retweets.

More specifically, the functions in \eqref{eq:fn.sep} are
assumed to take the following parametric forms,
\begin{equation}
  \label{eq:fn.par}
  \begin{split}
   p(\tau;\beta) &=e^{-\beta\tau},\\
   r(m;\gamma)   &=\gamma\log(m+1),\\
   \phi(t;\delta)&=\frac{\delta_2(\delta_1-1)}{\delta_1}
                   \left(1+\frac{\delta_2t}{\delta_1}\right)^{-\delta_1},
  \end{split}
\end{equation}
for parameters $\beta\geqslant 0,\gamma\geqslant 0,\delta_1>1$, and
$\delta_2>0$.  Here we have adopted an exponential decay form for the
infectivity function, based on the intuition that the infectivity, or
newsworthiness of a retweet should decay very quickly over time.  We
further assume that the impact function is linear in the number of
followers on a log scale, rather than on the original scale as in
\cite{Zhao2015}, because of the high degree of right skewness for the
distribution of the number of followers \citep{Cha2010, Kwak2010,
Bakshy2011}.  Our choice of the power law decay form for the memory
kernel is motivated by \cite{Zhao2015} and the
empirical findings of the heavy tailed distributions for the human
response time in social networks, reported in the literature
\citep{Barabasi2005,Crane2008,Zaman2014}.  Finally, as in
\cite{Zhao2015} and \cite{Kobayashi2016}, we also assume the event
marks $m_i$ are \iid\ with a common density function $f(\cdot)$
relative to a suitable reference measure on the space $\mathcal M$ of
event marks, and moreover, $m_i$ is independent of $\tau_i$ and
$\mathcal F_{\tau_i-}$ for all $i$.  As the excitation function
associated with an event is allowed to depend on the time of that
event, the model will be called a {\em Marked Self-Exciting Process
with Time-Dependent Excitation Function}, or \mas\ for short.

At this point we emphasize an important difference between the \mas\
we propose in this work and the \sei\ of \cite{Zhao2015}.  From
\eqref{eq:lmd}, we note that, unlike the \sei, the \mas\ has an
intensity process that is of a linear form similar to the
self-exciting process of \cite{Hawkes1971}, whose intensity
process takes the form,
\begin{equation*}
  \lambda(t)=\nu+\sum_{i=1}^{N(t-)}g(t-\tau_i).
\end{equation*}
In fact, if we choose $p(\tau)\equiv 1$ and $r(m)\equiv r$ for a constant
$r$ in \eqref{eq:fn.sep}, then \eqref{eq:lmd} reduces to the
time-varying version of the Hawkes process considered by
\cite{Chen2013,Chen2016}.  The linear structure of the intensity
process implies that the \mas\ can also be interpreted as a Poisson
cluster process, as the original Hawkes process or the generalized
version with a time-varying background intensity.  By this
interpretation, immigrants arrive according to a marked inhomogeneous
Poisson process with its intensity function equal to the baseline
intensity function $\nu(\cdot)$, and event mark distributed according to
the density function $f(\cdot)$.  Once an immigrant with mark $m$
arrives at $\tau$, it starts to independently produce children
according to a marked inhomogeneous Poisson process with intensity
function $\omega(\tau,m,\cdot)=p(\tau)r(m)\phi(\cdot)$ and event marks
distributed according to $f(\cdot)$, so that the total number of
children is Poisson distributed with mean
$\int_0^\infty\omega(\tau,m,t)\rmd t=p(\tau)r(m)$, and given the total
number of children, the waiting times to births of the children are
\iid\ with common density function $\phi(\cdot)$ if the order of
births is ignored.  Moreover, once an offspring of any generation is
born, say at time $\tau'$ and with mark $m'$, it starts to
independently produce children of its own according to a similar
marked inhomogeneous Poisson process with intensity function
$\omega(\tau',m',\cdot)$ and event marks distributed according to
$f(\cdot)$.  The events of the \mas\ process by time $t$ consist of all
immigrants and offspring of any generation that have arrived by time
$t$.  This Poisson cluster process interpretation implies an efficient
recursive cascading algorithm to simulate the \mas\ process as
described in Section~\ref{sec:pred}, which has important implications
for simulation based predictions by the \mas\ process.

Because of the Poisson cluster interpretation, the memory kernel
function $\phi(\cdot)$ in the \mas\ can also be called the {\em
offspring density function}, and the function $p(\cdot)r(\cdot)$
might be interpreted as the {\em branching ratio function} which
specifies how the branching ratio, that is, the average number of direct
(or generation 1) offspring from an individual (be it an immigrant
or an offspring), depends on the birth time and event mark of the
individual.  In contrast, the functions $p(\cdot)$ and $\phi(\cdot)$
in the \sei\ or the \tid\ model do not permit such a neat
interpretation.

It might also be of interest to note the difference between the
treatments of the background intensity in the \mas\ model and in the
Hawkes process model with a general time-varying background intensity.
In the former model, we require the baseline intensity function to be
proportional to the memory kernel $\phi(\cdot)$, while in the latter,
the background intensity and the memory kernel can take different
shapes. The advantage of our treatment is that it leads to a more
parsimonious model, while the time-varying background intensity
model can easily accommodate nonstationarity, such as that due to the
diurnal patterns of human activity levels.

\subsection{Parameter estimation}
\label{sec:par}
Before we can use the \mas\ model for future events prediction, we
need to first estimate the model parameters.  Since the event marks are
assumed to be \iid, their distribution can simply be estimated
by the empirical distribution of $m_i$, for $i=1,\dotsc,N(T)$.  The
main estimation problem is to estimate the parameter vector
$\theta=(\alpha,\beta,\gamma,\delta_1,\delta_2)\tr$.  To this end, we
shall use the maximum likelihood (ML) approach.  By the point process
theory \citep[][Proposition 7.3.III]{Daley2003}, the likelihood of the
\mas\ process based on observations over the interval $[0,T]$, where
$T$ denotes the censoring time, takes the following form
\begin{equation}
  \label{eq:lik}
  L(\theta)=\braces{\prod_{i=1}^{N(T)} \lambda(\tau_i)}
            \exp\brackets{-\int_0^T \lambda(t)\rmd t}
            \prod_{i=1}^{N(T)} f(m_i),
\end{equation}
where $\lambda(\cdot)$ depends on the parameters through
\eqref{eq:lmd}-\eqref{eq:fn.par}, and $f(\cdot)$ denotes the event
mark density, which is assumed to be free of the parameters $\theta$.

To compute the ML estimator of the parameter vector $\theta$ using
general-purpose numerical optimization routines, the efficient
evaluation of the likelihood function or its logarithm is very
important.  For this purpose, we need to be able to evaluate the
definite integral of the intensity function in \eqref{eq:lik}
efficiently.  Due to the linear structure of the intensity function,
the integral of the intensity function can be shown to take an
explicit form similar to the intensity function itself, and therefore
can be exactly computed without resorting to numerical quadrature
routines.  To show this, it is convenient to use the random measure
interpretation of a marked point process.  That is, we interpret
\begin{equation*}
  N(\rmd \tau, \rmd m)=\sum_{i=1}^\infty
  \delta_{(\tau_i,m_i)}(\rmd\tau,\rmd m)
\end{equation*}
as a random measure on $[0,\infty)\times \mathcal M$, so that
the intensity in \eqref{eq:lmd} can be written as
\begin{equation*}
  \lambda(t)= \nu(t)+\sum_{i=1}^{N(t-)}\omega(\tau_i,m_i,t-\tau_i) =
  \nu(t)+\int_{(0,t)\times \mathcal{M}}\omega(\tau,m,t-\tau)
  N(\rmd \tau,\rmd m).
\end{equation*}
Therefore, by Fubini's theorem, a change of variables, and the assumed
forms of the functions $\nu$, $\omega$ and $\phi$, we have
\begin{equation}
  \label{eq:Lmd}
  \begin{split}
    \int_0^T \lambda(t)\rmd t&=\int_0^T \nu(t)\rmd t
    + \int_0^T \int_{(0,t)\times \mathcal M}\omega(s,m,t- s)
    N(\rmd s,\rmd m)\rmd t\\
    &=\int_0^T  \nu(t)\rmd t  + \int_{(0,T)\times \mathcal M}
    \int_s^T \omega(s,m,t-s)\rmd t N(\rmd s,\rmd m)\\
    &=\int_0^T  \nu(t)\rmd t  + \int_{(0,T)\times \mathcal M}
    \int_0^{T-s} \omega(s,m,t)\rmd t N(\rmd s,\rmd m)\\
    &=\alpha \Phi(T)+ \sum_{i=1}^{N(T-)} p(\tau_i)r(m_i)\Phi(T-\tau_i),
  \end{split}
\end{equation}
where $\Phi(t)=\Phi(t;\delta)=\int_0^t \phi(s;\delta)\rmd s= 1-(1+\delta_2
t/\delta_1)^{-\delta_1+1}$, $t\geqslant 0$.

From the separable form of the likelihood function in \eqref{eq:lik} and
the assumption that the event mark distribution does not depend on the
parameter vector $\theta$, the ML estimation of the parameters
$\theta$ can be based on maximizing the logarithm of the part of the
likelihood that does not involve $f(\cdot)$, that is
\begin{equation*}
  \ell(\theta)=\sum_{i=1}^{N(T)}\log\lambda(\tau_i) - \int_0^T
  \lambda(t)\rmd t.
\end{equation*}
In practice, the maximization can be done numerically using various
general-purpose optimization routines.  In our numerical experiments,
we have used the downhill simplex method of \cite{Nelder1965}, which
is the default method used by the function \texttt{optim} in the
R software environment for statistical computing
\citep{Team2016}.

\subsection{Goodness-of-fit assessment}
\label{sec:gof}
The assessment of the goodness-of-fit of models to historical data
can guide us to seek models that can describe the observed data well
and therefore serves as the basis of predictions for future observations.
To assess the goodness-of-fit of the \mas, we shall use the residual
point process approach based on Papangelou's random time change
theorem \citep[][Theorem 7.4.I]{Daley2003}.  By the time change
theorem, with $\Lambda(t)=\int_0^t \lambda(s)\rmd s$ denoting the
cumulative intensity process, the transformed process
$N(\Lambda\inv(t))$ is a Poisson process with unit rate or
equivalently, the random times $\Lambda(\tau_i)$, $i=1,2,\dotsc$, will
be the event times of a unit rate Poisson process.  Therefore, if the
\mas\ with the parameters $\theta$ set to their ML estimates
$\hat\theta$ is a sufficient model for the observed event times up to
the censoring time $T$, then the transformed event times,
$\hat\Lambda(\tau_i)$, $i=1,\dotsc,N(T)$ should be approximately equal
in distribution to the event times of a unit rate Poisson process up
to time $\hat\Lambda(T)$, where $\hat\Lambda(t)$, $t>0$ is the plugin
estimate of the cumulative intensity
$\Lambda(t;\theta)=\int_0^t\lambda(s;\theta)\rmd s$, that is,
\begin{equation*}
  \hat\Lambda(t)=\Lambda(t;\hat\theta)=\hat\alpha \Phi(t;\hat\delta)+
  \sum_{i=1}^{N(t-)} p(\tau_i;\hat\beta)r(m_i;\hat\gamma)
  \Phi(t-\tau_i;\hat\delta),
\end{equation*}
with $p(\cdot)$ and $r(\cdot)$ defined in \eqref{eq:fn.par}, and
$\Phi(\cdot)$ defined as in \eqref{eq:Lmd}. As the conditional
distribution of the event times of a Poisson process in a fixed
interval given the total number of events in the interval is equal in
distribution to the order statistics of the same number of \iid\
random variables uniformly distributed in the interval, to assess the
goodness-of-fit of the \mas\ (or any point process model specified via
the intensity process), we can assess the uniformity of the
transformed event times $\hat\Lambda(\tau_i)$, $i=1,\dotsc,N(T)$, in
the interval $(0,\hat\Lambda(T)]$, informally using graphical
approaches such as the histogram or the Q-Q (quantile-quantile) plots,
or formally using statistical tests like the K-S (Kolmogorov-Smirnov)
test.  A similar residual analysis was performed by \cite{Ogata1988}
to assess the goodness-of-fit of point process models on earthquake
data.

\subsection{Predicting future number of events}
\label{sec:pred}
Given observations up to $T$, to predict the number of events from $T$
to a future time point $\T>T$, one commonly uses its conditional
expectation or its conditional median, which is optimal relative to
the mean squared error or the mean absolute error accordingly
\citep{Gneiting2011}.  To obtain the conditional expectation, we can
use either a solve-the-equation approach or a simulation based
approach.  The former approach involves deriving a functional equation
satisfied by the conditional expectation as a function of a future
time point, solving the equation, and evaluating the solution function
at the desired time point.  The latter approach involves simulating
the sample path of the \mas\ on the time interval $(T,\T]$ conditional
on the observations up to time $T$ for a large number of times,
counting the number of events on each simulated sample path, and using
the average of the simulated event counts to approximate its
expectation.  While the first approach is computationally less
expensive, the solution of the functional equation is not always easy
to obtain.  For the second approach, although it is relatively less
efficient, especially if the process to be simulated has a large
expected number of events, it is more robust than the first
approach.  To obtain the conditional median, the only option seems to
be a simulation based approach, which involves simulating the
conditional sample path of the \mas\ process a large number of times
and extracting the median of the resultant empirical distribution of
the number of events in the prediction interval.

Both solve-the-equation approach and simulation based approach
rely on the observation that, conditional on the history of the \mas\
process up to time $T$, its future evolution is the same as that of
another \mas\ process with a different baseline intensity function and
a similar excitation function.  Let $\tN(t)=N(T+t)-N(T)$, for
$t\geqslant 0$, $\tilde\tau_j=\tau_{N(T)+j}-T$, $\tilde
m_j=m_{N(T)+j}$ for $j=1,2,\dotsc$, and $\tilde{\mathcal
  F}_{t}=\mathcal F_{T+t}$, $t\geqslant 0$.  Then, the
$\tilde{\mathcal{F}}$-intensity process of $\tN(t)$ is given by
\begin{equation*}
  \begin{split}
    \tilde\lambda(t)=\lambda(T+t)={}&\nu(T+t) + \sum_{j=1}^{N(T)}
    \omega(\tau_j,m_j,T+t-\tau_j)
    +\sum_{j=N(T)+1}^{N(T+t -)} \omega(\tau_j,m_j,T+t-\tau_j)\\
    ={}&\tilde\nu(t) +
     \sum_{j=1}^{\tN(t-)}\tilde\omega(\tilde\tau_{j},\tilde
     m_j,t-\tilde \tau_j),
  \end{split}
\end{equation*}
where $\tilde\nu(\cdot)$ denotes the function
\begin{equation}
  \label{eq:nutilde}
  \tilde\nu(t)=\nu(T+t)+\sum_{j=1}^{N(T)}\omega(\tau_j,m_j,T+t-\tau_j),
\end{equation}
and $\tilde\omega(\cdot,\cdot,\cdot)$ denotes the function
\begin{equation}
  \label{eq:omegatilde}
  \tilde\omega(\tau,m,t)=\omega(T+\tau,m,t)=
  p(T+\tau)r(m)\phi(t)\equiv\tp(\tau)r(m)\phi(t).
\end{equation}
Therefore, $\tN(t)$, $t\geqslant 0$ is a \mas\ process with baseline
intensity function $\tilde\nu$ and excitation function $\tilde\omega$
given as above in~\eqref{eq:nutilde} and~\eqref{eq:omegatilde}
respectively.  The excitation function $\tilde\omega$ has a similar
separable form as $\omega$, with $r$ and $\phi$ the same as before,
and infectivity function $\tp$ equal to a time shift of the
previous infectivity function, that is,  $\tilde p(\tau)=p(T+\tau)$.

To calculate the expected number of events
$\smallE{\smallrightbar{N(\T)-N(T)}\mathcal F_T}$ without resorting to
simulations, we first note from the definition of the conditional
intensity that,
\begin{equation}
  \label{eq:predpt}
  \begin{split}
    &\E{\rightbar{N(\T)-N(T)}\mathcal F_T}=
     \E{\rightbar{\tN(\T-T)}\mathcal F_T}\\
    ={}&\E{\rightbar{\int_0^{\T-T}\tilde\lambda(s)\rmd s}\mathcal F_T}
    =\int_0^{\T-T}\E{\rightbar{\tilde\lambda(s)}\mathcal F_T}\rmd s
    =\int_0^{\T-T}\blmd(s) \rmd s,
  \end{split}
\end{equation}
with $\blmd(s)=\smallE{\smallrightbar{\tilde\lambda(s)}\mathcal F_T}$
denoting the mean intensity function of $\tN(t)$ given $\mathcal
F_{T}$.  By the independence between event marks and previous
event times, we have
\begin{align*}
\blmd(t)={}&\E{\rightbar{\tilde\lambda(t)}\mathcal F_T}\\
    ={}&\E{\rightbar{\tilde\nu(t) + \int_{(0,t)\times \mathcal M}
    \tilde \omega(\tau,m,t-\tau) \tilde N (\rmd \tau, \rmd m) }
    \mathcal F_T}\\ \stepcounter{equation}\tag{\theequation}\label{eq:blmd}
    ={}&\E{\rightbar{\tilde\nu(t) + \int_{(0,t)\times \mathcal M}
    \tp(\tau)r(m)\phi(t-\tau)\tilde\lambda(\tau)\rmd \tau \rmd F(m)}
    \mathcal F_T}\\ 
    ={}&\tilde\nu(t) + \int_{\mathcal M}r(m)\rmd F(m) \int_0^t
    \tp(\tau)\phi(t-\tau)\E{\rightbar{\tilde\lambda(\tau)}\mathcal F_T}
    \rmd \tau\\
    ={}&\tilde\nu(t) + R\int_0^t \tp(\tau)\phi(t-\tau)\blmd(\tau)
    \rmd \tau,
\end{align*}
where we have also used $\tN(\rmd \tau,\rmd m)$ to denote the
associated random measure again, and $F$ denotes the distribution of the
\iid\ event marks, while $R=\E{r(m_i)}=\int_{\mathcal M} r(m)\rmd F(m)$
is the expected total excitation effect due to an event.

In general, we need to solve the integral equation in \eqref{eq:blmd}
numerically to obtain $\blmd(t)$ on $[0,\T-T]$ and use it in finding
the conditional expectation of the number of events in
\eqref{eq:predpt}.  One method to solve \eqref{eq:blmd} is to
approximate $\blmd(t)$ by a flexible parametric function and
identify the parameters by requiring both sides of the equation
to be equal or approximately equal at sufficiently many points
in the interval $[0,\T-T]$.  Examples of the flexible
parametric functions to approximate $\blmd(t)$ include a B-spline
function with a specified order and knot sequence, or a truncated
Fourier series.  In both cases, the unknown parameters of the
approximating function can be obtained by solving a linear equation of
the unknown parameters.  In practice, we would try approximating
functions with increasing flexibility until convergence in the
solution is achieved.  We select the B-spline function in this work as
a method to find $\blmd(t)$ for its ease of implementation and
computational stability.  Specifically, we let
$B(t)=(B_1(t),\dotsc,B_k(t))\tr$ denote the set of B-spline basis
functions of a certain order on the interval $(0,\T-T]$, and assume
that $\blmd(t)\approx B(t)\tr \eta$ for a $k$-vector $\eta$.  Plugging
this into \eqref{eq:blmd}, we have the following equation
of $\eta$,
\begin{equation}
  \label{eq:eta}
  \begin{split}
    &B(t)\tr\eta =\tilde\nu(t) + \braces{R\int_0^t
    \tp(\tau)\phi({t-\tau})B(\tau)\tr\eta\rmd \tau}.
  \end{split}
\end{equation}
To solve \eqref{eq:eta} for $\eta$, we evaluate both sides of
\eqref{eq:eta} at sufficiently many ($\geqslant k$) $t$ values in the
interval $(0,\T-T]$, and solve the resulting overdetermined linear
system using the method of least squares to get $\eta$.  Once $\eta$ is
obtained, the predicted value is calculated as
\begin{equation*}
  \braces{N(\T)-N(T)}_{\mathrm{pred}}=
  \brackets{\int_{0}^{\T-T}B(t) \rmd t} \tr \eta.
\end{equation*}
In evaluating the integrals in \eqref{eq:eta}, we often need to use
numerical quadrature routines.  In our case, we have used
the R function \texttt{integrate} for this purpose.

To simulate the \mas\ process $\tN$ over the interval $(0,\T-T]$, we
can use the following cascading algorithm, which is a generalization
of that used for the simulation of non-stationary self-exciting point
processes \citep{Chen2013,Chen2016}.  A similar algorithm has been used
by \cite{Chen2017} to simulate renewal Hawkes processes.
\begin{enumerate}
\item Simulate an inhomogeneous Poisson process $N^0$ with time-varying
  intensity function $\tilde\nu(t)$ on $(0,\T-T]$ and denote the event
  times of $N^0$ by $\tau^0_j$, $j=1,\dotsc,N^0(\T-T)$.
\item Generate the associated event marks $m^0_j$ independently
  from the event mark distribution $F$ and call the events
  $(\tau^0_j,m^0_j), j=1, \dotsc, N^0(\T-T)$ generation 0
  events.
\item For each generation $0$ event $(\tau^0_j,m^0_j)$,
  simulate an inhomogeneous marked Poisson process $N^1_j$,
  with intensity function $\tilde \omega(\tau^0_j,m^0_j,\cdot)$
  and event mark distribution $F$, on the interval $(0,\T-T
  -\tau^0_j]$ and denote the corresponding events by
  $(\tau^1_{jk},m^1_{jk}), k=1,\dotsc,N^1_{j}(\T -T-\tau^0_j)$.
  The collection of events $\{(\tau^0_j
    +\tau^1_{jk},m^1_{jk}); k= 1,\dotsc,N^1_j(\T -T-\tau^0_j),
    j=1,\dotsc,N^0(\T-T)\}$ are referred to as generation 1 events.
\item Continue generating events of generations 2, 3, $\dotsc$
  similarly on intervals of decreasing lengths, until a
  generation has no events.
\item The events of all generations are pooled together to form
  the collection of all events of the \mas\ $\tN$ process on
  the interval $(0,\T-T]$.
\end{enumerate}
The algorithm shown above requires the simulation of inhomogeneous
Poisson processes, which can be achieved using the thinning algorithm
of \cite{Lewis1979}.  In our numerical experiments, we have used the R
implementation \texttt{simPois} from the \texttt{IHSEP} package.  Our
implementation of the above cascading algorithm is based on a simple
modification of the function \texttt{simHawkes1} from the R package
\texttt{IHSEP}.

To predict the number of events in the interval $(T,\T]$, we simulate
the sample path of the process $\tN(t)$ over the interval $(0,\T-T]$ for
a large number (say 100) of times, and count the number of events on each
simulated sample path.  The mean or median of these simulated event numbers
will then be our point prediction of the number of events of the \mas\ process $N$
in the interval $(T,\T]$.
In practice, when we use the fitted model to make predictions, whether
by using the solve-the-equation approach or by using the simulation based
approach, the unknown functions $\tilde\nu$ and $\tilde\omega$, and
the event mark distribution $F$ need to be replaced by their
respective estimators.  In our numerical experiments, we use the plugin
estimators $\tilde\nu(t;\hat\theta)$ and
$\tilde\omega(\cdot,\cdot,\cdot;\hat\theta)$ for $\tilde\nu$ and
$\tilde\omega$, and the empirical distribution function $\hat F$ of
the event marks $m_1,\dotsc,m_{N(T)}$ for $F$.  One implication is that
the constant $R$ in \eqref{eq:blmd} is set to $\hat R=
\int_{\mathcal{M}}r(m)\rmd \hat F(m)=\sum_{i=1}^{N(T)} r(m_i) /N(T)$.
Finally, we note that, if the target of prediction is the total number
of events of the process $N$ in the interval $(0,\T]$, then we simply
add the observed number in the interval $(0,T]$, that is, $N(T)$,
to the predicted number in the interval $(T,\T]$.

\section{Application to the tweet data}
\label{SEC:res}
In this section, we report the results of applying the proposed
model and inference methodologies to the tweet data.  The performance of
our prediction methods is also compared to those of the \sei\
and the \tid\ model.

\subsection{The model fit}
\label{sec:mdfit}
We fitted the \mas\ model to the 71,815 retweet cascades in the
training data set described in Section~\ref{sec:data} with different
censoring times, using the maximum likelihood method described in
Section~\ref{sec:par}. The estimated parameter values with censoring
time of seven days are highly skewed, with the median estimates of
$\alpha$, $\beta$, $\gamma$, $\delta_1$, and $\delta_2$ equal to
$48.349$, $0.072$, $7.209$, $1.416$, and $0.007$ respectively. To have
some idea about the typical parameter values found in practice, we
display in Table~\ref{tab:est} the estimated parameter values for
the five sample cascades shown in Figure~\ref{fig:smpdat}, together
with their final popularity. 

\begin{table}[hbt]
  \centering
  \caption{Fitted parameter values on the sample cascades shown in
    Figure~\ref{fig:smpdat}}
  \label{tab:est}
  \begin{tabular}{ccccccc}
    \hline
    Sample Cascade&
    $\hat\alpha$&$\hat\beta$&$\hat\gamma$&$\hat\delta_1$&$\hat\delta_2$&$N(\T)$\\\hline
    1 &  5.711 & 0.024 & 1.455 & 1.254 & 0.173 & 159\\
    2 &  3.075 & 0.021 & 6.351 & 1.414 & 0.029 &  85\\
    3 & 58.136 & 0.246 & 1.144 & 1.490 & 0.001 &  55\\
    4 &  8.209 & 0.031 & 2.095 & 1.444 & 0.040 &  74\\
    5 &  4.173 & 0.019 & 5.049 & 1.229 & 0.046 &  89\\\hline
  \end{tabular}
\end{table}

The estimated values of the parameter $\beta$ suggest very fast 
decays of infectivity, with the times taken for the infectivity 
to drop to 1\% of the initial levels vary from about 19 seconds 
($\log(100)/0.246$
$=18.7$ seconds) in sample cascade 3 to about four minutes 
($\log(100)/0.019=242.4$ seconds) in sample cascade 5. 
While the estimated values of the shape parameter of the memory
kernel $\delta_1$ are more or less similar to each other, the scale
parameter $\delta_2$ is substantially more variable. In particular,
the extremely small $\hat\delta_2$ value of $0.001$ in sample cascade
3 implies a very long range memory effect, which, together with a
relatively large $\hat\beta$ value, suggest that the later retweets
are more likely to be generated by the original tweet or retweets
within the first few seconds of the original tweet (if any), while in
sample cascade 1, where the $\hat\delta_2$ value is $0.173$, the later
retweets are more likely to be generated by more recent retweets. The
estimated values of the scale parameter $\alpha$ of the baseline
intensity function, together with the values of the $\delta$
parameters and the final popularity, suggest highly variable
proportions of generation 0 retweets, ranging from 3.4\%
($=5.711\Phi(\T;1.254,0.173)/159$) in sample cascade 1 to nearly 100\%
($=58.136\Phi(\T;1.490,0.001)/55$) in sample cascade 3. Also, the 
estimated $\gamma$ values on the five sample cascades have quite
substantial variation, with the increase in the excitation effect
associated with one unit increase in the number of followers (on the
log scale) of a retweeting account vary from 1.144 to 6.351 units.

By the goodness-of-fit assessment method described in
Section~\ref{sec:gof}, we tested the uniformity of the point process
residuals $\hat\Lambda(\tau_i)$ over the interval $(0,\hat\Lambda(T)]$
using the K-S test.  At significance levels of 0.01 and 0.05 with
different censoring times, the percentages of the 71,815 cascades
where the estimated \mas\ model passes the residual uniformity test
are shown in Table~\ref{tab:fitness}. 

\begin{table}[H]
  \centering
  \caption{The percentages of cascades in the training data set where
           the \mas\ model passes the goodness-of-fit test at different
           significance levels and censoring times}
  \label{tab:fitness}
  \begin{tabular}{cccccccc}\hline
  \multirow{2}{*}{Significance level} &
  \multicolumn{7}{c}{Censoring time (hours)} \\
  \cline{2-8}
  &   2    &    4    &    6    &    8    &   10    &   12    &  168\\
  \hline
   $0.01$&92.0\% & 88.2\% & 85.8\% & 84.2\% & 82.8\% & 81.8\% & 74.9\%\\
   $0.05$&89.3\% & 84.7\% & 81.9\% & 80.1\% & 78.5\% & 77.5\% & 69.2\%\\
    \hline
  \end{tabular}
\end{table}

From this table we note that the percentage of cascades from
which the estimated model passes the test decreases when the censoring
time increases.  This is to be expected as the amount of data
increases with the censoring time, implying that the difficulty of
finding a fitting model also increases.  At significance level of
0.01, when fitted to the complete cascade data, that is, with the
censoring time of 168 hours (seven days), the \mas\ model passes the
goodness-of-fit test on roughly 75\% of the cascades.  By the
censoring time of 12 hours, the \mas\ model passes the goodness-of-fit
test on the majority (approximately 82\%) of the cascades.  Given that
the majority of the retweets, or 80\% on average, have already
happened within 12 hours of the posting of the original tweets, we
conclude that the \mas\ model is able to describe the retweeting
dynamics reasonably well.

\subsection{Popularity prediction}
\label{sec:res}
For each of the 94,254 tweets in the test data set, we applied the
fitted \mas\ model with the retweet cascades censored at different
times to predict their final popularity, using the prediction methods
discussed in Section~\ref{sec:pred}.  For the purpose of comparison, we
also obtained the predictions based on the \sei\ of \cite{Zhao2015}
and the \tid\ model of \cite{Kobayashi2016}.  We only report the
results of comparison with these two methods, because they were found
to outperform other methods in the literature, such as those reported
in \cite{Crane2008}, \cite{Agarwal2009}, \cite{Szabo2010}, and
\cite{Gao2015}, both in our own numerical experiments and in the works
of \cite{Zhao2015} and \cite{Kobayashi2016}.

Our point prediction of the ``final'' popularity of a tweet, or the
total number of retweets by time $\T=7\ \mathrm{ days}$, using the
\mas\ model estimated with the retweet cascade observed up to the
censoring time $T$, is given by $N(\T)_{\mathrm{pred}}=N(T)+
(N(\T)-N(T))_{\mathrm{pred}}$, where $(N(\T)-N(T))_{\mathrm{pred}}$ is
obtained either as the conditional expectation using the
solve-the-equation approach or the simulation based approach, or as
the conditional median using the simulation based approach.  Our
numerical experiments have confirmed that in the case of conditional
expectation, the two approaches produce identical predictions up to a
negligible numerical error, as expected.  When using the simulation
based approach, we note that, for some very popular tweets, the
retweeting cascades are very long and the numbers of retweeting events
to be simulated in the prediction intervals are very large, and
therefore simulations can take a long time to complete.  This issue
was also noted by \cite{Kobayashi2016}.  A trick we used to mitigate
this issue is to simulate the process $\tN$ with a smaller baseline
intensity function, say $\tilde\nu(\cdot)/S$ with $S = 100$ or larger,
and inflate the simulated event numbers by the factor $S$.  For the
majority of the cascades, a moderately large number of simulation
replications, such as 100 or even 50, was enough to produce a
prediction consistent with that by the solve-the-equation approach.
The same set of simulations were used to calculate the median based
prediction.

To assess the performance of the conditional mean based predictions,
we first follow the recent literature \citep{Zhao2015,Kobayashi2016}
and use the Absolute Percentage Error (APE),
\begin{equation*}
  APE = \left|\frac{N(\T)_{\mathrm{pred}}-N(\T)}{N(\T)}\right|,
\end{equation*}
to compare the accuracy of prediction by different models.  Each
prediction method under evaluation was applied to each of the 94,254
retweet cascades in the test data set with censoring times
$T=2,4,\dotsc,12$ hours as most retweeting events would have occurred
within the first few hours, based on our analysis on the training data
set.  For each censoring time, we calculated the APEs of the
conditional mean predictions by the proposed model and by the two
competing models.  The predictions by the \sei\ approach were
calculated using the R package \texttt{seismic}.  The predictions by
the \tid\ model approach were calculated using the algorithm described
in \cite{Kobayashi2016} with the window size parameter
$\Delta_{\mathrm{obs}}$ \citep[cf.][p. 194]{Kobayashi2016} in the
estimation step set to one hour.  Due to the lack of a principled
approach to select the window size, we chose this value based on
experimenting with different values and selecting the one that
produced reasonable estimates of the infectivity function by visual
inspection.

\begin{figure}[hbt]
  \centering
  \includegraphics[width=0.75\textwidth,height=0.475\textwidth,keepaspectratio]{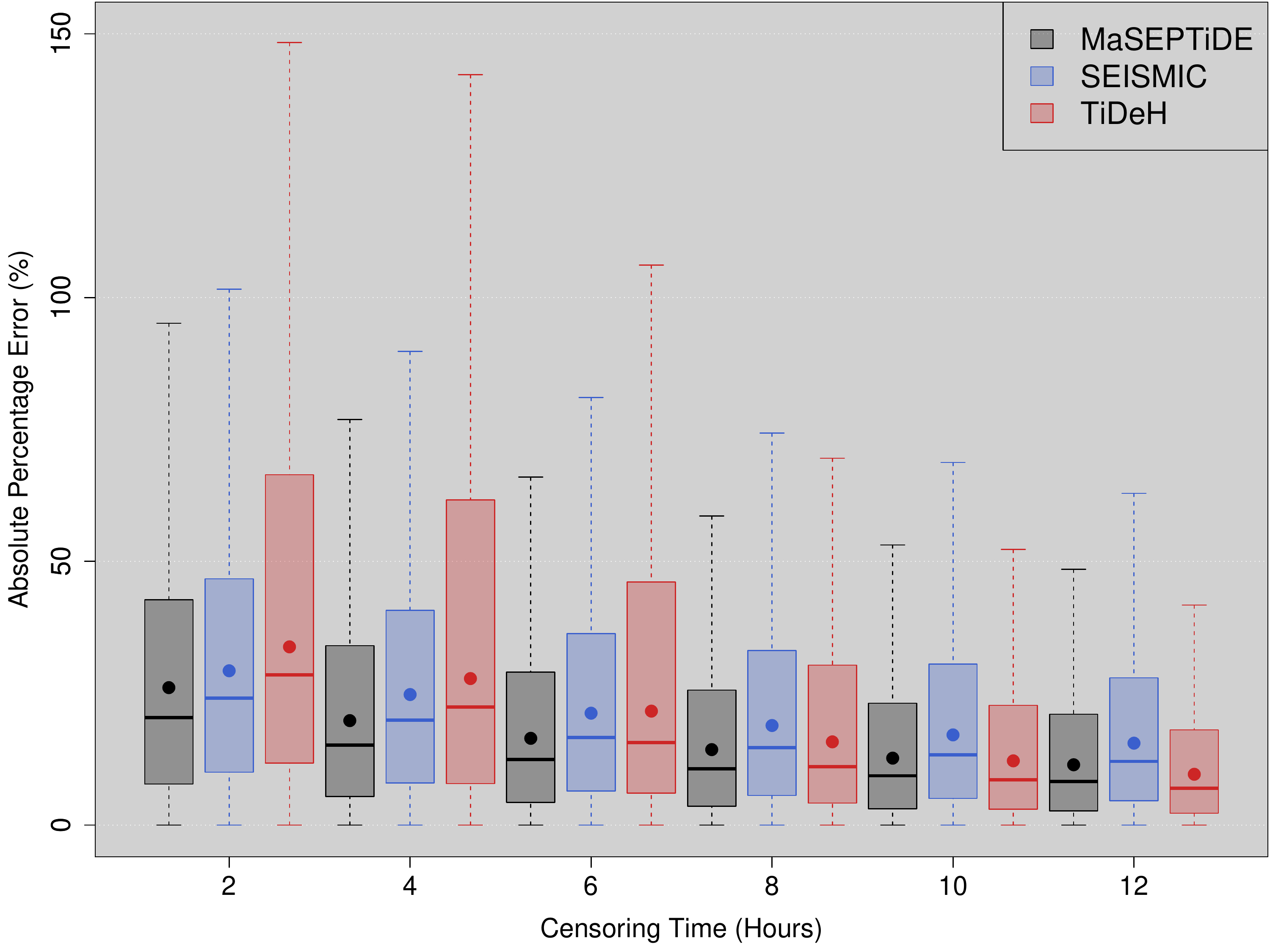}
  \caption{Boxplots of the Absolute Percentage Errors (APEs) of
    predictions by the \mas\ model, the \sei\ and the \tid\
    model, for censoring times $T=2,4,\dotsc,12$ hours.
    The horizontal thick bar in each boxplot indicates
    the median while the circular point indicates the
    respective mean of APEs.  Both the mean and median
    of APEs demonstrate the superior performance and
    stability of the \mas\ model.}
  \label{fig:perf}
\end{figure}

The boxplots of the APEs of conditional mean predictions by the three
models (\mas, \sei, and \tid) with different censoring times are shown
in Figure~\ref{fig:perf}.  In each boxplot, the horizontal thick bar
indicates the median APE, and the circular point indicates the mean APE.
The actual values of the median and mean APE are given
in Table~\ref{tab:ape}. 

\noindent From Figure~\ref{fig:perf} and Table~\ref{tab:ape},
the \mas\ prediction has consistently smaller median APE and mean
APE for each $T$ than the \sei\ prediction.  Compared to the \tid\
prediction, the \mas\ prediction has clearly better performances
when $T=2,4,6$ hours, either by the median APE or by the mean
APE.  The performances of these two models are comparable when $T=8$
hours, but the \mas\ is at a slight disadvantage when $T=10,12$
hours.

\begin{table}[H]
  \centering
  \caption{Median APEs and mean APEs of the popularity predictions by
    different approaches with observations up to various censoring times $T$}
  \label{tab:ape}
  \begin{tabular}{c|ccc|ccc}
    \hline $T$&
    \multicolumn{3}{c|}{Median APE (\%)} &
    \multicolumn{3}{c}{Mean APE (\%)}
    \\ \cline{2-7} (hours) &
    \mas\ & \sei\ & \tid\ &
    \mas\ & \sei\ & \tid\
    \\
    \hline
    2  & 19.1 & 22.8 & 23.7 & 26.1 & 29.3 & 33.8 \\
    4  & 13.9 & 18.6 & 17.1 & 19.8 & 24.8 & 27.8 \\
    6  & 11.2 & 15.1 & 12.7 & 16.5 & 21.2 & 21.6 \\
    8  &  9.5 & 13.1 &  9.3 & 14.3 & 18.9 & 15.8 \\
    10 &  8.2 & 11.7 &  7.4 & 12.7 & 17.1 & 12.2 \\
    12 &  7.3 & 10.6 &  5.9 & 11.4 & 15.5 &  9.6 \\
    \hline
  \end{tabular}
\end{table}

\begin{figure}[hbt]
  \centering
  \includegraphics[width=0.475\textwidth,height=0.8\textwidth,
  keepaspectratio]{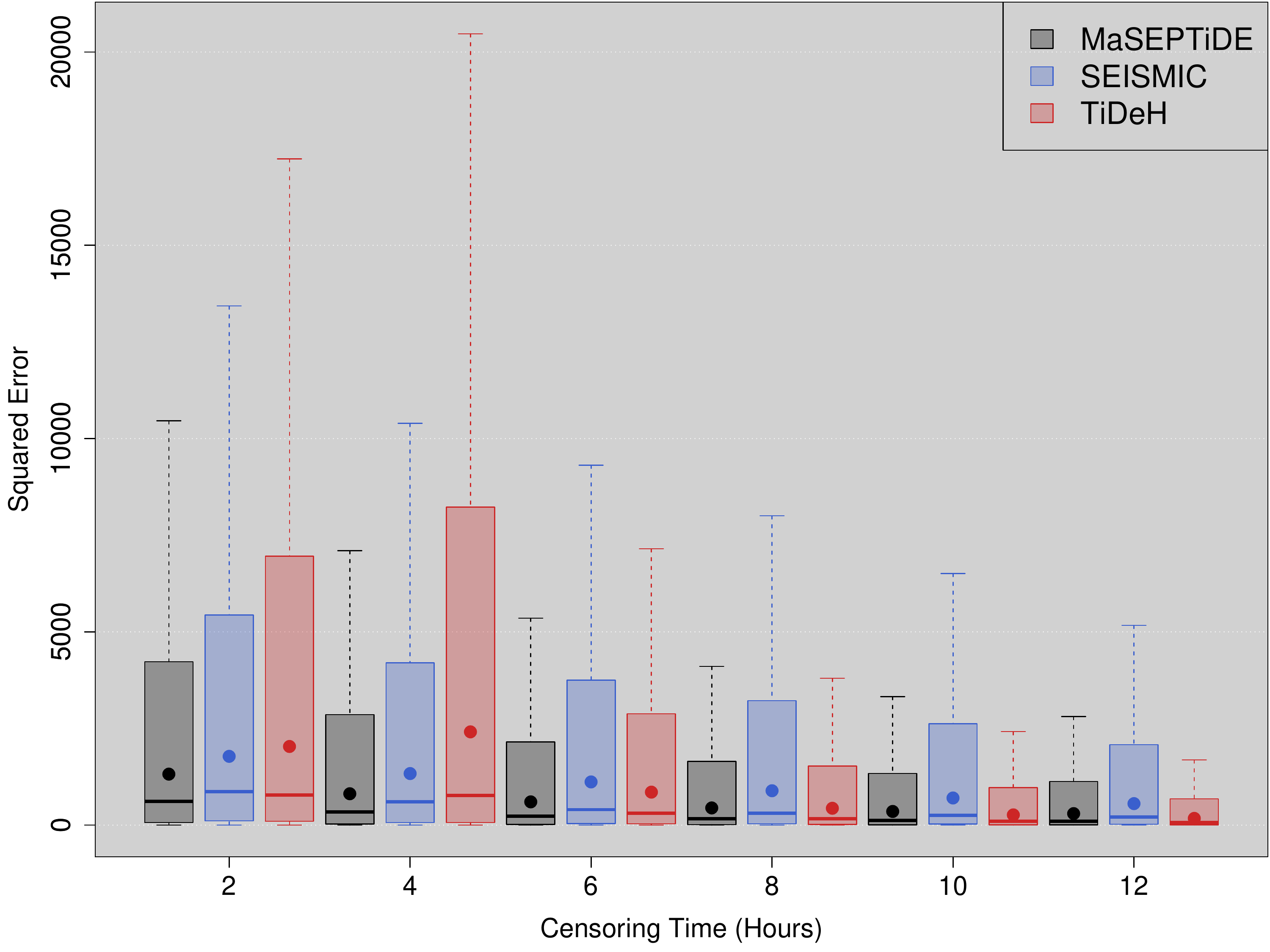}
  \includegraphics[width=0.475\textwidth,height=0.8\textwidth,
  keepaspectratio]{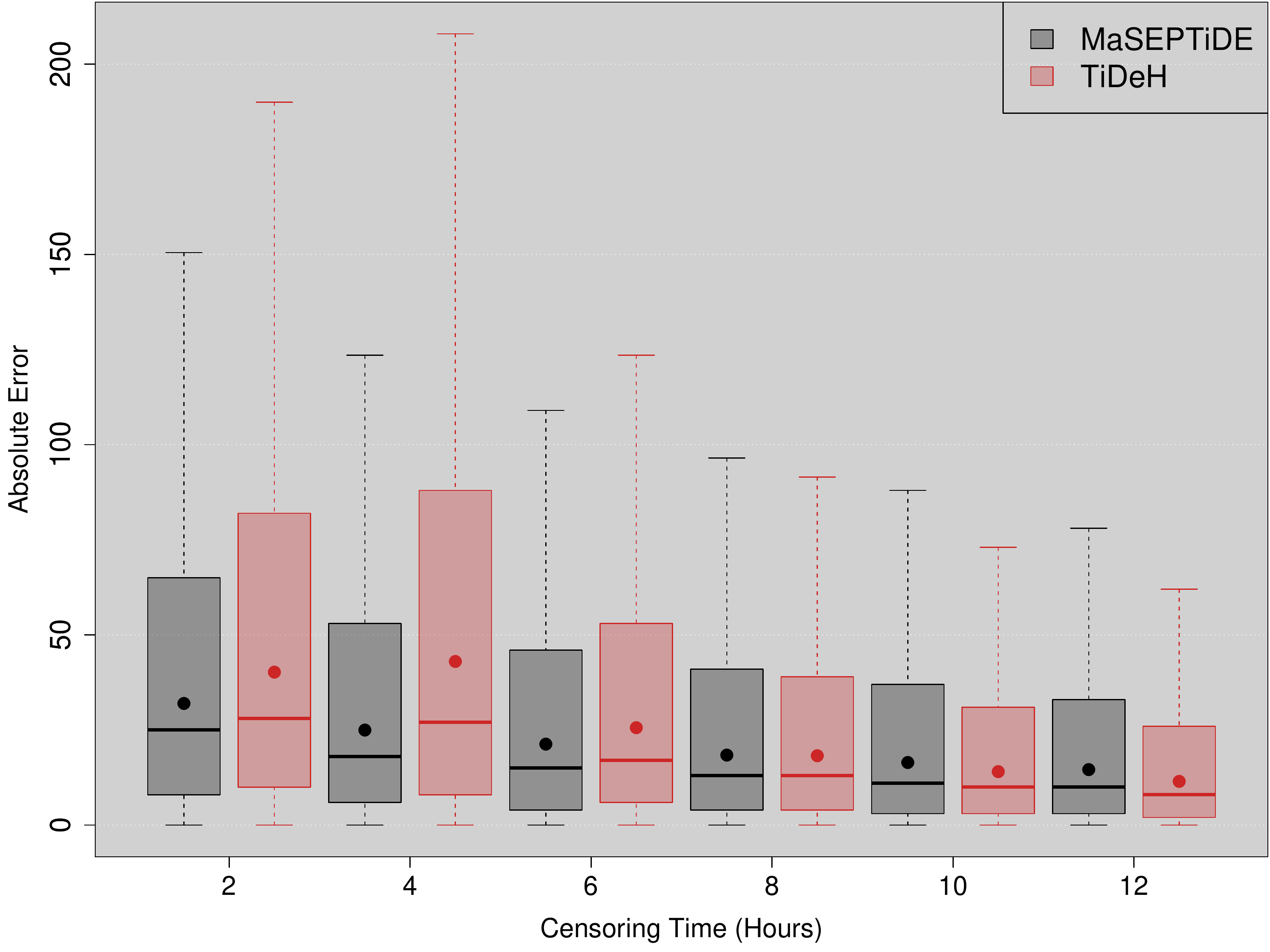}
  \caption{Left: squared prediction errors when the mean of the
    predictive distribution is used as the point prediction; Right:
    absolute prediction errors when the median is used.  The thick
    horizontal bar in each boxplot shows the median of the errors, and
    the circular point shows the mean of the errors. }
  \label{fig:se&ae}
\end{figure}

Despite its widespread use in evaluating prediction performance, the
APE as a prediction error measure is not consistent with the feature
of the predictive distribution used as a point prediction here, which
is the expectation/mean; see \cite{Gneiting2011} for a systematic
discussion of this issue.  A more appropriate prediction error measure
when the conditional mean is used as the point prediction is the
squared error.  Therefore, we also calculated the squared errors of the
predictions by the three models.  The boxplots of the squared errors by
the three models at different censoring times are shown in the left
panel of Figure~\ref{fig:se&ae}.

As the models can occasionally produce extremely large predictions,
and even infinity in the case of \sei, the outlying values were not
shown in the boxplot for better visualization.  The mean squared
prediction errors (MSEs) at different censoring times are indicated by
the circular points in the boxplots, and their squared roots, that is,
the root mean squared errors (RMSEs), are shown in Table~\ref{tab:rmse-mae}.  
From Figure~\ref{fig:se&ae} and Table~\ref{tab:rmse-mae}, 
we note that, using the RMSE as the performance measure, 
the \mas\ prediction outperforms the \sei\ prediction at all 
the censoring times, and similar to the conclusion drawn based 
on the median APEs, the \mas\ again, outperforms the \tid\
model when $T=2,4,6$ hours but slightly underperforms when
$T = 8,10,12$ hours.  In comparison, the \sei\ only
outperforms the \tid\ model when $T=2,4$ hours.

\begin{table}[hbt]
\centering
\caption{Root mean squared errors (RMSEs) and mean absolute errors (MAEs) 
of predictions at different censoring times}
\label{tab:rmse-mae}
\begin{tabular}{c|ccc|cc}
\hline
$T$ &
\multicolumn{3}{c|}{RMSE} &
\multicolumn{2}{c}{MAE}
\\ \cline{2-6} (hours)&
\mas & \sei & \tid &
\mas & \tid \\
\hline
2   & 36.3 & 42.2 & 45.1 & 32.0 & 40.2\\
4   & 28.5 & 36.5 & 49.1 & 25.0 & 43.0\\
6   & 24.5 & 33.4 & 29.2 & 21.3 & 25.6\\
8   & 21.1 & 29.8 & 20.9 & 18.4 & 18.2\\
10  & 18.8 & 26.5 & 16.3 & 16.5 & 14.1\\
12  & 17.3 & 23.6 & 13.2 & 14.6 & 11.5 \\\hline
\end{tabular}
\end{table}

To compare the prediction performances when using the conditional
median by different models, we use the mean absolute error (MAE) as
the criterion of comparison, as advised by \cite{Gneiting2011}.  The
conditional median prediction by the \mas\ model was calculated by the
simulation based approach described in Section~\ref{sec:pred}.  The
conditional median prediction by the \tid\ model was similarly
calculated using a simulation based approach, although the simulation
of the \tid\ model was achieved by a less efficient method where the
events have to be simulated serially one after another, using the
rejective method of \cite{Lewis1979}.  The conditional median
prediction by the \sei\ is not included in this comparison because
this model does not specify the form of the intensity process beyond
the censoring time, and therefore we cannot calculate the conditional
median using the simulation based approach.  The right panel of
Figure~\ref{fig:se&ae} shows the absolute errors of the conditional
median predictions by the \mas\ and the \tid\ models at different
censoring times, where, as before, the circular points indicate the
MAEs of the predictions at the corresponding censoring times.  See
also Table~\ref{tab:rmse-mae} for MAE values.  By the MAE, the
conditional median prediction by the \mas\ model is clearly superior
to that by the \tid\ model at the censoring times $T=2,4,6$
hours, and is comparable albeit slightly inferior at the larger
censoring times $T=8,10,12$ hours.

By all the performance evaluation criteria considered, the prediction
by the \mas\ model is clearly more accurate than by the two competing
models, especially when prediction needs to be made based on shorter
observation times, for example, within six hours or shorter of the 
posting of the original tweet.

\section{Conclusion and discussion}
\label{SEC:con}
In this work, we have proposed a marked self-exciting point process
model, called the \mas, to model the retweeting dynamics and to predict
tweet popularity.  The \mas\ is able to model a large number
of retweet cascades adequately, and its prediction performance
is superior to those of the competing models and approaches in the
literature that require the same input.

When the prediction is based on observing a cascade for a long period
of time, the approach based on the \tid\ model by \cite{Kobayashi2016}
is found to outperform our model by a small margin.  However,
considering the fact that this small advantage of the \tid\ model is
not realized until the retweet cascades are observed for eight hours
or longer, when the majority of the retweets would have already
happened, its practical significance is rather limited.  On the
contrary, the approach based on the \mas\ is able to provide accurate
predictions of the final popularity based on observations within two 
hours of posting of the original tweet.  Another issue with the \tid\
model is that the nonparametric estimation step to obtain the initial
raw estimate of the infectivity curve needs a large amount of data to
work well.  In fact, in their numerical experimentation,
\cite{Kobayashi2016} only verified the superior performance of their
prediction approach relative to the \sei\ on 738 very long cascades
(containing 2,000 or more retweets), which account for less than 0.5\%
of all the cascades.  In contrast, the approach based on the \mas\ is
fully parametric and does not require as much data to
estimate.

The specific parametric forms of the functions in the \mas\ model have
been selected from a class of candidate models by comparing their
goodness-of-fit on the retweet cascades in the training data set and
identifying the model that can fit most of the cascades.  In the class
of candidate models, we have considered other parametric forms of
the component functions, such as infectivity functions that decay at
polynomial rate, and memory kernel functions that decay exponentially
fast.  The model with the specific forms of the component functions
reported herein has the best goodness-of-fit on the training data.

To further improve the \mas\ model, more complex models, such as those
that incorporate the calendar time effects \citep{Fox2016,Kobayashi2016} are
worth considering.  Another aspect of our approach that can be improved is
that our approach still requires the observation of the retweet cascade for
a substantial amount of time to accumulate enough data to identify the model,
even though the required observation time is much less compared to
approaches based on other models such as the \tid\ model.  If we make
stronger assumptions on the model parameters across the cascades, then
parameter estimation might be achieved using only training data, which
will allow us to predict its final popularity as soon as a tweet
is published, or even before it is published.

Finally, an important limitation of the data considered in our work,
originally collected by \cite{Zhao2015}, is that the data contains
only cascades with at least 49 retweets.  Such data
is by no means representative of all tweets published by Twitter users,
as vast majority of the tweets do not get even a single
retweet.  Therefore, models developed based on such data are only
useful for popularity predictions of reasonably popular
tweets.  To develop models suitable for the predictions of the
popularity of average tweets, one would need to collect suitable
random samples of tweets and their retweet cascades, and build
models accordingly.\vspace{12pt}

\noindent{\textbf{Acknowledgements.}} The authors thank Qingyuan Zhao
and Ryota Kobayashi for the clarifications on the implementation
details of their popularity prediction methods.  This research is sponsored by 
the Ministry of Higher Education, Malaysia, and includes
computations using the Linux computational cluster Katana supported by
the Faculty of Science, UNSW Australia.

\bibliographystyle{imsart-nameyear}
\def\bibfont{\small}
\bibliography{MaSEPTiDE_ArXiV}

\end{document}